\documentclass[runningheads]{llncs}
\usepackage{graphicx}
\usepackage{hyperref}

\usepackage[utf8x]{inputenc}
\usepackage{amsmath}
\usepackage{amssymb}
\usepackage{upgreek}
\usepackage{xcolor}
\usepackage{colortbl}
\usepackage{enumitem}
\usepackage{cite}

\usepackage{tikz}
\usetikzlibrary{arrows,positioning,calc,shapes.geometric}
\usepackage{tabularx}

\definecolor{keywordcolor}{rgb}{0.7, 0.1, 0.1}   
\definecolor{tacticcolor}{rgb}{0.0, 0.1, 0.3}    
\definecolor{commentcolor}{rgb}{0.4, 0.4, 0.4}   
\definecolor{stringcolor}{rgb}{0.5, 0.3, 0.2}    
\definecolor{symbolcolor}{rgb}{0.1, 0.2, 0.7}    
\definecolor{sortcolor}{rgb}{0.1, 0.5, 0.1}      
\definecolor{attributecolor}{rgb}{0.7, 0.1, 0.1} 
\definecolor{errorcolor}{rgb}{1, 0, 0}           

\usepackage{listings}
\usepackage{flushend}
\usepackage{xspace}
\usepackage{underscore}

\newcommand{\mathlib}{\textsf{mathlib}\xspace}

\newcommand{\lean}[1]{\lstinline[language=lean]{#1}}

\usepackage{scalefnt}

\usepackage{setspace}

\begin{document}
\title{Maintaining a Library of Formal Mathematics
\thanks{The first author is supported by the Sloan Foundation (grant G-2018-10067).
The second and third authors receive support
from the European Research Council (ERC) under
the European Union's Horizon 2020 research and innovation
program (grant agreement No. 713999, Matryoshka)
and from the Dutch Research Council (NWO) under the
Vidi program (project No. 016.Vidi.189.037, Lean Forward).}
}
%
%
\author{Floris van Doorn\inst{1}\orcidID{0000-0003-2899-8565} \and \\
Gabriel Ebner\inst{2}\orcidID{0000-0003-4057-9574} \and \\
Robert Y.\ Lewis\inst{2}\orcidID{0000-0002-5266-1121}}
\authorrunning{F.\ van Doorn, G.\ Ebner, and R.Y.\ Lewis.}
%
\institute{University of Pittsburgh, Pittsburgh PA 15260, USA \\
\email{fpvdoorn@gmail.com} \\
\and
Vrije Universiteit Amsterdam, 1081HV Amsterdam, The Netherlands \\
\email{gebner@gebner.org, r.y.lewis@vu.nl}}
\maketitle              
\begin{abstract}
The Lean mathematical library \textsf{mathlib}
is developed by a community of users
with very different backgrounds and levels of experience.
To lower the barrier of entry for contributors
and to lessen the burden of reviewing contributions,
we have developed a number of tools for the library
which check proof developments for subtle mistakes in the code
and generate documentation suited for our varied audience.

\end{abstract}

\section{Introduction}
\label{section:intro}

As a tool for managing mathematical knowledge,
a proof assistant offers many assurances.
Once a result has been formalized,
readers can confidently believe that
the relevant definitions are fully specified,
the theorem is stated correctly,
and there are no logical gaps in the proof.
A body of mathematical knowledge,
represented by formal definitions and proofs
in a single theorem proving environment,
can be trusted to be coherent.

Logical coherence, however,
is only one of many properties
that one could wish of a mathematical corpus.
The ideal corpus can be modified, extended, and queried
by users who do not have expert knowledge of the entire corpus or the underlying system.
Proof assistant libraries do not always fare so well in this respect.
Most of the large mathematical libraries in existence
are maintained by expert users
with a significant time cost.
While external contributions are easily checked for logical consistency,
it typically takes manual review to check that contributions cohere with the system in other ways---e.g.,
that lemmas are correctly marked for use with a simplification tactic.
It can be difficult or impossible for outsiders
to understand the library well enough to contribute themselves.

The \mathlib library \cite{mathlib20} is a corpus
of formal mathematics, programming, and tactics
in the Lean proof assistant \cite{Mour15a}
that is managed and cultivated by a community of users.
The community encourages contributions from novice users,
and the rapid growth of the library has threatened to overwhelm its appointed maintainers.
The maintenance difficulty is compounded by
the library's extensive use of type classes and context-dependent tactics.
Misuse of these features is not always easy to spot,
but can lead to headaches in later developments.

To ease the burdens on new users and maintainers alike,
we have incorporated into \mathlib tools for
checking meta-logical properties of declarations
and collecting, generating, and displaying documentation in an accessible way.
The use of these tools has already had a large impact on the community.
We aim here to explain the goals and design principles of these tools.
While some details are specific to Lean and \mathlib,
we believe that these considerations apply broadly
to libraries of formal mathematical knowledge.

\section{Lean and \mathlib}
\label{section:lean}

Lean offers a powerful metaprogramming framework
that allows Lean programs to access the system's syntax and core components \cite{EURAM17}.
All of the linting tools described in Sect.~\ref{section:linting}
are implemented in Lean,
without the need for external plugins or dependencies.
They are distributed as part of the \mathlib library.

Lean metaprograms are frequently used to implement \emph{tactics},
which transform the proof state of a declaration in progress.
They can also implement top-level \emph{commands},
which interact with an environment outside the context of a proof.
Examples include
\lean{#find}, which searches for declarations matching a pattern,
and \lean{mk_simp_attribute}, which defines a new collection of simplification lemmas.
\emph{Transient} commands like \lean{#find},
which do not modify the environment,
customarily start with \lean{#}.
Tactics and commands interact with the Lean environment and proof state
through the \lean{tactic} monad,
which handles side effects and failure conditions
in a purely functional way. 
Finally, Lean supports tagging declarations with \emph{attributes}
as a way to store metadata.
Within the \lean{tactic} monad,
metaprograms can access the list of declarations tagged with a certain attribute.

The \mathlib project is run by a community of users
and encourages contributions from people with various backgrounds.
The community includes many domain experts,
people with expert knowledge of the mathematics being formalized
but who are less familiar with the intricacies of the proof assistant.
Linting and documentation are useful
for every user of every programming language,
but are especially helpful for such domain experts,
since they often work on deep and intricate implementations
without a broad view of the library.

An example of this is seen in \mathlib's \emph{structure hierarchy}~\cite[Sect.\ 4]{mathlib20}.
The library extensively uses type classes
to allow definitions and proofs to be stated
at the appropriate level of generality
without duplication.
Type classes are a powerful tool,
but seemingly innocent anti-patterns in their use
can lead to unstable and unusable developments.
Even experienced users find it difficult to avoid these patterns,
and they easily slip through manual code review.

The \mathlib library and community are growing at a fast pace.
As of May 15, 2020, 
the library contains over 170,000 lines of non-whitespace, non-comment code,
representing a 25\% increase over five months,
and 42,000 
declarations, excluding internal and automatically generated ones,
a 23\% increase.
Contributions have been made by 85 people,
a 16\% increase over the same time period.
264 commits were made to the \mathlib git repository in April 2020;
while a small number were automatically generated,
each commit typically corresponds to a single approved pull request.
We display more statistics about the project's growth on the community website.\footnote{
  \url{https://leanprover-community.github.io/mathlib_stats.html}
}
The library covers a wide range of subject matter,
enough to serve as a base for numerous projects
that have formalized complex and recent mathematical topics~\cite{BCM20,Dahm19,Han20}.

\section{Semantic Linting}
\label{section:linting}

Static program analysis,
the act of analyzing computer code without running the code,
is widely used in many programming languages.
An example of this is \emph{linting},
where source code is analyzed to flag faulty or suspicious code.
Linters warn the user about various issues,
such as syntax errors,
the use of undeclared variables,
calls to deprecated functions,
spacing and formatting conventions,
and dangerous language features.

In typed languages like Lean,
some of these errors are caught by the elaborator or type checker.
The system will raise an error
if a proof or program has a different type than the declared type
or if a variable is used that has not been introduced.
However, other problems can still be present
in developments that have been accepted by Lean.
It is also possible that there are problems with the metadata of a declaration,
such as its attributes or documentation.
These mistakes are often not obvious at the time of writing a declaration,
but will manifest at a later time.
For example, an instance might be declared that will never fire,
or is likely to cause the type class inference procedure to loop.

We have implemented a package of semantic linters in \mathlib
to flag these kinds of mistakes.
These linters are \emph{semantic} in the sense that
they take as input a fully elaborated declaration and its metadata.
This is in contrast to a \emph{syntactic} linter,
which takes as input the source code as plain text.
The use of semantic linters allows us
to automatically check for many commonly made mistakes,
using the abstract syntax tree (the elaborated term in Lean's type theory) for the type or value of a declaration.
Syntactic linters would allow for testing of
e.g.\ the formatting of the source code,
but would not help with many of the tests we want to perform.

The linters can be  used
to check one particular file or all files in \mathlib.
Running the command \lean{#lint} at any point in a file
prints all the linter errors up to that line.
The command \lean{#lint_mathlib} tests all imported declarations in \mathlib.
Occasionally a declaration may be permitted to fail a lint test,
for example, if it takes an unused argument to satisfy a more general interface.
Such lemmas are tagged with the attribute \lean{@[nolint]},
which takes a list of tests that the declaration is allowed to fail.
The continuous integration (CI) workflow of \mathlib
automatically runs the linters on all of \mathlib
for every pull request made to the library.



For some of the mistakes detected by our linters, 
it is reasonable to ask 
whether they should even be allowed by the system in the first place.
The core Lean tool aims to be small, permissive, and customizable;
enforcing our linter rules at the system level
would cut against this philosophy.
Projects other than \mathlib may choose to follow different conventions,
or may be small enough
to ignore problems that hinder scalability.
Stricter rules, of course, can create obstacles to finishing a project.
By incorporating our checks into our library instead of the core Lean system,
we make them available to all projects that depend on \mathlib 
without forcing users to comply with them.

\subsection{Linter Interface}
A linter is a wrapper around a metaprogram with type
\lean{declaration → tactic} \lean{(option string)}. 
Given an input declaration \lean{d},
the test function returns \lean{none} if \lean{d} passes the test
and \lean{some error_msg} if it fails.
These test functions work within the \lean{tactic} monad
in order to access the elaborator and environment,
although some are purely functional
and none modify the environment.
The type \lean{linter} bundles such a test function with formatting strings.

The package of linters is easily extended:
a user simply defines and tags a declaration of type \lean{linter}.
In Fig.~\ref{fig:doc-blame}
one sees the full definition of the \lean{doc_blame} linter,
described in Sect.~\ref{subsection:simple-linters}.

\begin{figure}[t]
\begin{lstlisting}
/-- Reports definitions and constants that are missing doc strings -/
meta def doc_blame_report_defn : declaration → tactic (option string)
| (declaration.defn n _ _ _ _ _) := doc_string n >> return none <|> return "def missing doc string"
| (declaration.cnst n _ _ _) := doc_string n >> return none <|> return "constant missing doc string"
| _ := return none

/-- A linter for checking definition doc strings -/
@[linter, priority 1450] meta def linter.doc_blame : linter :=
{ test := λ d, mcond (bnot <$> has_attribute' `instance d.to_name) (doc_blame_report_defn d) (return none),
  no_errors_found := "No definitions are missing documentation.",
  errors_found := "DEFINITIONS ARE MISSING DOCUMENTATION STRINGS" }
\end{lstlisting}
\caption{A linter that tests whether a declaration has a documentation string.}
\label{fig:doc-blame}
\end{figure}

We have focused on implementing these linters
with actionable warning messages.
Since the errors they detect are often subtle
and can seem mysterious to novice users,
we try to report as clearly as possible
what should change in a declaration
in order to fix the warning.

\subsection{Simple Linters}
\label{subsection:simple-linters}

A first selection of \mathlib linters
checks for simple mistakes
commonly made when declaring definition and theorems.

\paragraph{Duplicated namespaces.}
Declaration names in Lean are hierarchical,
and it is typical to build an interface for a declaration
in its corresponding namespace.
For example, functions about the type \lean{list}
have names such as \lean{list.reverse} and \lean{list.sort}.
Lean's \lean{namespace} sectioning command
inserts these prefixes automatically.
However, users often write a lemma with a full name
and then copy it inside the namespace.
This creates identifiers like \lean{list.list.reverse};
it can be difficult to notice the duplication without careful review.
The \lean{dup_namespace} linter flags declarations
whose names contain repeated components.

\paragraph{Definitions vs.\ theorems.}
Lean has separate declaration kinds for definitions and theorems.
The subtle differences relate to byte code generation
and parallel elaboration.
It is nearly always the case that
a declaration should be declared as a theorem
if and only if its type is a proposition.
Because there are rare exceptions to this, the system does not enforce it.
The \lean{def_lemma} linter checks for this correspondence,
so that the user must explicitly approve any exceptions.

\paragraph{Illegal constants.}
The Lean core library defines \lean{a > b} to be \lean{b < a},
and similarly for \lean{a ≥ b}.
These statements are convertible,
but some automation, including the simplifier,
operates only with respect to syntactic equality.
For this reason, it is convenient to pick a normal form for equivalent expressions.
In \mathlib, we prefer theorems to be stated in terms of \lean{<} instead of \lean{>}.
The \lean{ge_or_gt} linter checks that the disfavored constants
do not appear in the types of declarations.

\paragraph{Unused arguments.}
A very common beginner mistake
is to declare unnecessary arguments to a definition or theorem.
Lean's useful mechanisms for auto-inserting parameters in namespaces and sections
can unfortunately contribute to this.
The \lean{unused_arguments} linter checks that each argument to a declaration
appears in either a subsequent argument or the declaration type or body.

\paragraph{Missing documentation.}
The \mathlib documentation guidelines require
every definition to have a doc string (Sect.~\ref{section:docs}).
Since doc strings are accessible by metaprograms,
we are able to enforce this property with a linter,
called \lean{doc_blame} (Fig.~\ref{fig:doc-blame}).
Missing doc strings are the most common linter error caught in CI.

\subsection{Type Class Linters}
\label{subsection:typeclass}

Lean and \mathlib make extensive use of \emph{type classes}~\cite{Wadl89}
for polymorphic declarations.
Of the 42,000 declarations in \mathlib,
465 are type classes and 4600 are type class instances.
In particular, type classes are used to manage the hierarchy of mathematical structures.
Their use allows definitions and theorems to be stated at high levels of generality
and then applied in specific cases without extra effort.
Arguments to a declaration are marked as \emph{instance implicit}
by surrounding them with square brackets.
When this declaration is applied,
Lean runs a depth-first backward search through its database of instances
to satisfy the argument.
Type classes are a powerful tool,
but users often find the underlying algorithms opaque,
and their misuse can lead to performance issues~\cite{Sels20}.
A collection of linters
aims to warn users about this misuse.

\paragraph{Guiding type class resolution.}

Instances can be assigned a positive integer \emph{priority}.
During type class resolution the instances with a higher priority are tried first.
Priorities are optional,
and in \mathlib most instances are given the default priority.
Assigning priorities optimally is difficult.
On the one hand, we want to try instances that are used more frequently first,
since they are most likely to be applicable.
On the other hand, we want to try instances that fail more quickly first,
so that the depth-first search does not waste time on unnecessary searches.

While we cannot automatically determine the optimal priority of instances,
there is one class of instances we want to apply last,
namely the \emph{forgetful instances}.
A forgetful instance is an instance that applies to every goal,
like the instance \lstinline{comm_group α → group α},
which forgets that a commutative group is commutative.
Read backward as in the type class inference search,
this instance says that to inhabit \lstinline{group α}
it suffices to inhabit \lstinline{comm_group α}.

Forgetful instances contrast with \emph{structural instances}
such as \lstinline{comm_group α} \lstinline{→} \lstinline{comm_group β →} \lstinline{comm_group} \lstinline{(α × β)}. 
We want to apply structural instances before forgetful instances,
because if the conclusion of a structural instance unifies with the goal,
it is almost always the desired instance.
This is not the case for forgetful instances,
which are always applicable,
even if the extra structure or properties are not available for the type in question.
In this case, the type class inference algorithm will do an exhaustive search
of the new instance problem, which can take a long time to fail.
The \lean{instance_priority} linter
enforces that all forgetful instances have priority below the default.

Another potential problem with type class inference
is the introduction of metavariables in the instance search.
Consider the following definition of an $R$-module type class.
\begin{lstlisting}
class module (R : Type u) (M : Type v) :=
(to_ring : ring R)
(to_add_comm_group : add_comm_group M)
(to_has_scalar : has_scalar R M)
/- some propositional fields omitted -/
\end{lstlisting}
If we make the projection \lean{module.to_ring} an instance,
we have an instance of the form \lstinline{module R M → ring R}.
This means that during type class inference,
whenever we search for the instance \lstinline{ring α},
we will apply \lean{module.to_ring}
and then search for the instance \lstinline{module α ?m}, where \lean{?m} is a metavariable.
This type class problem is likely to loop,
since most \lean{module} instances will apply in the case that the second argument is a variable.

To avoid this,
in \mathlib the type of \lean{module} actually takes as arguments
the ring structure on \lean{R} and the group structure on \lean{M}.
The declaration of module looks more like this:
\begin{lstlisting}
class module (R : Type u) (M : Type v) [ring R] [add_comm_group M] :=
(to_has_scalar : has_scalar R M)
/- some propositional fields omitted -/
\end{lstlisting}
Using this definition, there is no instance from modules to rings.
Instead, the ring structure of \lean{R} is carried as an argument
to the module structure on \lean{M}.
The \lean{dangerous_instance} raises a warning
whenever an instance causes a new type class problem
that has a metavariable argument.

\paragraph{Misused instances and arguments.}
Misunderstanding the details of type class inference
can cause users to write instances that can never be applied.
As an example, consider the theorem which says that
given a continuous ring homomorphism $f$ between uniform spaces,
the lift of $f$ to the completion of its domain is also a ring homomorphism.
The predicate \lean{is_ring_hom f} is a type class in \mathlib,
and this theorem was originally written as a type class instance:
\begin{lstlisting}
is_ring_hom f → continuous f → is_ring_hom (completion.map f)
\end{lstlisting}
However, \lean{continuous f} is not a type class,
and this argument does not appear in the codomain \lean{is_ring_hom (completion.map f)}.
There is no way for the type class resolution mechanism to infer this argument
and thus this instance will never be applied.
The \lean{impossible_instance} linter checks declarations for this pattern,
warning if a non-type class argument does not appear
elsewhere in the type of the declaration.

A dual mistake to the one above is
to mark an argument as instance implicit
even though its type is not a type class.
Since there will be no type class instances of this type,
such an argument will never be inferable.
The \lean{incorrect_type_class_argument} linter checks for this.
While the linter is very simple,
it checks for a mistake that is difficult to catch in manual review,
since it requires complete knowledge of the \mathlib instance database.

\paragraph{Missing and incorrect instances.}
Most theorems in \mathlib are type-polymorphic,
but many hold only on \emph{inhabited} types.
(Readers used to HOL-based systems should note that
Lean's type theory permits empty types,
e.g.\ an inductive type with no constructors.)
Inhabitedness is given by a type class argument,
so in order to apply these theorems,
the library must contain
many instances of the \lean{inhabited} type class.
The \lean{has_inhabited_instance} linter checks,
for each concrete \lean{Type}-valued declaration,
that conditions are given to derive
that the type is inhabited.

The \lean{inhabited} type class is itself \lean{Type}-valued.
One can computably obtain a witness \lean{t : T}
from an instance of \lean{inhabited T};
it is possible to have multiple distinct (nonconvertible) instances of \lean{inhabited T}.
Sometimes the former property is not necessary,
and sometimes the latter property can create problems.
For instance, instances deriving \lean{inhabited T}
from \lean{has_zero T} and \lean{has_one T}
would lead to non-commuting diamonds
in the type class hierarchy.
To avoid this, \mathlib defines a weaker type class,
\lean{nonempty}, which is \lean{Prop}-valued.
Lean propositions are \emph{proof-irrelevant},
meaning that any two terms of the same \lean{Prop}-valued type are indistinguishable.
Thus \lean{nonempty} does not lead to non-commuting diamonds,
and is safe to use in situations where \lean{inhabited} instances would cause trouble.

The \lean{inhabited_nonempty} linter checks for declarations
with \lean{inhabited} arguments that can be weakened to \lean{nonempty}.
Suppose that a \lean{Prop}-valued declaration
takes an argument \lean{h : inhabited T}.
Since Lean uses dependent types,
\lean{h} may appear elsewhere in the type of the declaration.
If it doesn't,
it can be weakened to \lean{nonempty T},
since the elimination principles are equivalent for \lean{Prop}-valued targets.
Weakening this argument makes the declaration more widely applicable.

\subsection{Linters for Simplification Lemmas}
\label{subsection:simp-linters}

Lean contains a \lean{simp} tactic for (conditional) term rewriting.
Similar tactics, such as Isabelle's \texttt{simp}~\cite{Nipkow02},
are found in other proof assistants.
Users can tag theorems using the \lean{@[simp]} attribute.
The theorems tagged with this attribute
are collectively called the \emph{simp set}.
The \lean{simp} tactic uses lemmas from the simp set,
optionally with extra user-provided lemmas,
to rewrite until it can no longer progress.
We say that such a fully simplified expression
is in \emph{simp-normal form} with respect to the given simp set.

\begin{figure}[t]
\begin{lstlisting}
@[simp] lemma zero_add (x : ℕ) : 0 + x = x := /- ... -/

example (x : ℕ) : 0 + (0 + x) = x := by simp
\end{lstlisting}
\caption{Example usage of the simplifier.
}
\label{figure:simp-example}
\end{figure}

The simplifier is used widely:
\mathlib contains over 7000 simp lemmas, 
and the string \lean{by simp} occurs almost 5000 times,
counting only a small fraction of its invocations.
However, care needs to be taken when formulating simp lemmas.
For example,
if both \lean{a = b} and \lean{b = a} are added as simp lemmas,
then the simplifier will loop. 
Other mistakes are more subtle.
We have integrated several linters
that aid in declaring effective simp lemmas.

\paragraph{Redundant simplification lemmas.}
We call a simp lemma redundant
if the simplifier will never use it for rewriting.
This redundancy property depends on the whole simp set:
a simp lemma is not redundant by itself,
but due to other simp lemmas that
break or subsume it.
One way a simp lemma can be redundant is
if its left-hand side is not in simp-normal form.

Simplification proceeds from the inside out,
starting with the arguments of a function before simplifying the enclosing term.
Given a term \lean{f (0 + a)},
Lean will first simplify \lean{a},
then it will simplify \lean{0 + a} to \lean{a}
using the simp lemma \lean{zero_add} (Fig.~\ref{figure:simp-example}),
and then finally simplify \lean{f a}.

A lemma stating \lean{f (0 + x) = g x} will never be used by the simplifier:
the left-hand side \lean{f (0 + x)} contains the subterm \lean{0 + x}
which is not in simp-normal form.
Whenever the simplifier tries to use this lemma to rewrite a term,
the arguments to \lean{+} have already been simplified,
so this subterm can never match.

It is often not immediately clear
whether a term is in simp-normal form.
The first version of the \lean{simp_nf} linter only checked
that the arguments of the left-hand side of a simp lemma
are in simp-normal form.
This first version identified more than one hundred lemmas across \mathlib
violating this condition.
In some cases, the lemma satisfied this condition in the file where it was declared,
but later files contained simp lemmas that simplified the left-hand side.

Simp lemmas can also be redundant if one simp lemma generalizes another simp lemma.
The simplifier always picks the \emph{last} simp lemma
that matches the current term.
(It is possible to override this order
using the \lean{@[priority]} attribute.)
If a simp lemma is followed by a more general version,
then the first lemma will never be used,
such as \lean{length_singleton} in the following example.
It is easy to miss this issue at first glance
since \lean{[x]} and \lean{x::xs} look very different,
but \lean{[x]} is actually parsed as \lean{x::[]}.
\begin{lstlisting}
@[simp] lemma length_singleton : length [x] = 1 := rfl
@[simp] lemma length_cons : length (x::xs) = length xs + 1 := rfl
\end{lstlisting}

Both of these issues are checked by the \lean{simp_nf} linter.
It runs the simplifier on
the left-hand side of the simp lemma,
and examines the proof term returned by the simplifier.
If the proof of the simplification of the left-hand side uses
the simp lemma itself,
then the simp lemma is not redundant.
In addition, we also assume that the simp lemma is not redundant
if the left-hand side does not simplify at all,
as is the case for conditional simp lemmas.
Otherwise the linter outputs a warning
including the list of the simp lemmas
that were used.

\paragraph{Commutativity lemmas.}
Beyond conditional term rewriting,
Lean's simplifier also has limited support for
ordered rewriting with commutativity lemmas
such as \lean{x + y = y + x}.
Naively applying such lemmas clearly leads to non-termination,
so the simplifier only uses these lemmas
if the result is smaller as measured by a total order on Lean terms.
Rewriting with commutativity lemmas results in nice normal forms
for expressions without nested applications of the commutative operation.
For example, it reliably solves the goal \lean{f (m + n) = f (n + m)}.
However, in the presence of nested applications,
the results are unpredictable:
\begin{lstlisting}
example (a b : ℤ) : (a + b) + -a = b := by simp /- works -/
example (a b : ℤ) : a + (b + -a) = b := by simp /- fails -/
\end{lstlisting}
The \lean{simp\_comm} linter checks that
the simp set contains no commutativity lemmas.

\paragraph{Variables as head symbols.}
Due to the implementation of Lean's simplifier,
there are some restrictions on simp lemmas.
One restriction is that
the head symbol of the left hand side of a simp lemma must not be a variable.
For example, in the hypothetical (conditional) lemma
\begin{lstlisting}
∀ f, is_homomorphism f → f (x + y) = f x + f y
\end{lstlisting}
the left-hand side has head symbol \lean{f},
which is a bound variable,
and therefore the simplifier will not rewrite with this lemma.
The \lean{simp_var_head} linter ensures that
no such lemmas are accidentally added to the simp set.



\section{Documentation}
\label{section:docs}

Programming language documentation serves very different purposes for different audiences,
and proof assistant library documentation is no different.
When creating documentation for Lean and \mathlib, we must address users who
\begin{itemize}
 \item are new to Lean and unfamiliar with its syntax and paradigms;
 \item would like an overview of the contents of the library;
 \item would like to understand the design choices made in an existing theory;
 \item would like a quick reference to the interface for an existing theory;
 \item need to update existing theories to adjust to refactorings or updates;
 \item would like to learn to design and implement tactics or metaprograms; and
 \item would like a quick reference to the metaprogramming interface.
\end{itemize}
Many of these goals are best served with user manuals or tutorials~\cite{Avig14}.
Such documents are invaluable,
but there is a high cost to maintaining and updating them.
They are most appropriate for material that does not often change,
such as the core system syntax and logical foundations.
From the perspective of library maintenance,
we are particularly interested in \emph{internal documentation},
that is, documentation which is directly written in the \mathlib source files.
Since the library evolves very quickly,
it is essential
to automatically generate as much of the reference material as possible.
Furthermore, human-written text should be close to what it describes,
to make it harder for the description and implementation to diverge.

We focus here on a few forms of this internal documentation.
\emph{Module documentation}, written at the top of a \mathlib source file, is intended to
describe the theory developed in that file,
justify its design decisions,
and explain how to use it in further developments.
(A Lean source file is also called a \emph{module}.)
\emph{Declaration doc strings} are written immediately before definitions and theorems.
They describe the behavior or content of their subject declarations.
In supported editors, these doc strings are
automatically displayed when the cursor hovers over a reference to the declaration.
\emph{Decentralized documentation}
is not localized to a particular line or file of the library,
although it may originate in a certain place;
it is expected to be collected and displayed post hoc.
An example of this is tactic documentation:
\mathlib defines hundreds of interactive tactics in dozens of files,
but users expect to browse them all on a single manual page.

Some features of proof assistants
(and of Lean and \mathlib in particular)
encourage a different style of documentation
from traditional programming languages.
Since Lean propositions are proof-irrelevant,
only the statement of a theorem,
not its proof term,
can affect future declarations.
Thus theorems are self-documenting in a certain sense:
the statement of a theorem gives a complete account of its content,
in contrast to a definition of type \lstinline{ℕ → ℕ}, for example.
We require doc strings on all \mathlib definitions
but allow them to be omitted from theorems.
While it is often helpful
to have the theorem restated or explained in natural language,
the manual burden of writing and maintaining these strings
for the large amount of simple lemmas in \mathlib
outweighs the gain of the natural language restatement.
Nonetheless, doc strings are strongly encouraged on important theorems
and results with nonstandard statements or names.

\subsection{Generation Pipeline}

In the style of many popular programming languages,
we generate and publish HTML documentation
covering the contents of \mathlib.
The generation is part of \mathlib's continuous integration setup.

Perhaps unusually for this kind of tool,
our generator does not examine the \mathlib source files.
Instead, it builds a Lean environment that imports the entire library
and traverses it using a metaprogram.
The metaprogramming interface allows access to the file name, line number, and doc string
for any particular declaration,
along with module doc strings.
By processing a complete environment
we can display terms using notation declared later in the library,
and include automatically generated declarations
that do not appear in the source.
We can also associate global information with declarations:
for example, we can display a list of instances for each type class.

The generation metaprogram produces a JSON file
that contains all information needed
to print the module, declaration, and decentralized documentation.
A separate script processes this database
into a searchable HTML website.\footnote{\url{https://leanprover-community.github.io/mathlib_docs/}}

\subsection{Declaration Display}

\begin{figure}[t]
 \includegraphics[width=\textwidth]{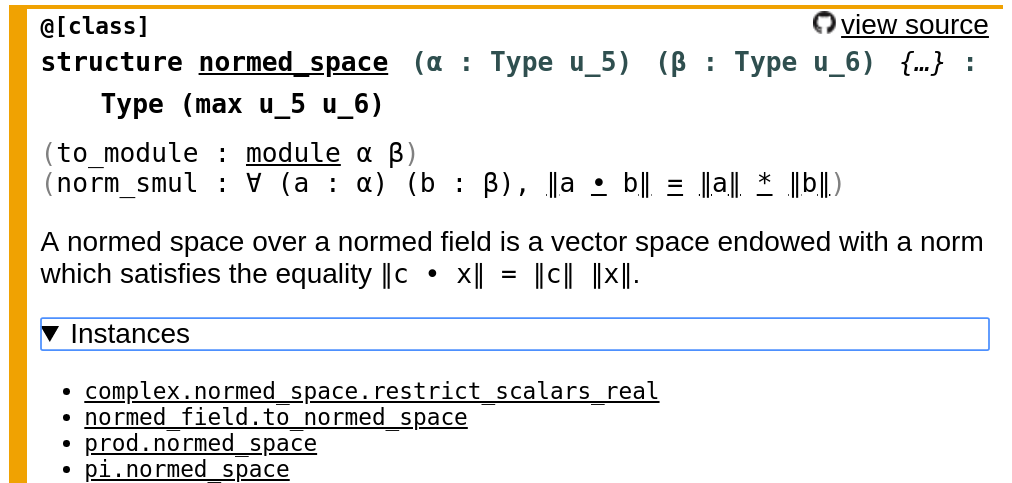}
 \caption{The generated documentation entry for the \lean{normed_space} type class.
   The implicit arguments can be expanded by clicking on \lean{\{...\}}.}
 \label{figure:normedspace}
\end{figure}

The majority of the documentation is oriented around modules.
For each Lean source file in \mathlib,
we create a single HTML page displaying
the module documentation
and information for each declaration in that file.
Declarations appear in the same order as in the source,
with an alphabetical index in a side panel.
For each declaration, we print various pieces of information (Fig.~\ref{figure:normedspace}).

The declaration name is printed
including its full namespace prefix.
Lean declarations have four possible kinds:
\lean{theorem}, \lean{definition}, \lean{axiom}, and \lean{constant}.
We print the declaration kind
and use it to color the border of the entry for a visual cue.
The type of the declaration is printed
with implicit arguments hidden by default.
This gives an easy reference as to how the declaration can be applied.
Each type can be expanded to display all arguments.
When a declaration has a doc string,
it is displayed beneath the type.

Lean represents the type former and constructors of an inductive type as separate constants.
We display them together, mirroring the Lean syntax for an inductive definition.
Similarly, we print the constructor and fields of a structure mirroring the input syntax.

We do not display all of the attributes applied to a declaration,
but show those in a predefined list, including \lean{simp} and \lean{class}.
For declarations tagged as type classes,
we display a collapsible list of instances of this class
that appear elsewhere in the library.
For definitions,
we display a collapsible list of the equational lemmas
that describe their associated reduction rules.
We also link to the exact location
where the declaration is defined in the source code.

We believe that this display achieves many of our design goals.
The module documentation provides
an overview of a particular theory for newcomers
and general implementation details for experts.
The declaration display serves as an API reference,
displaying information concisely
with more details readily available.
The same framework works to document both
the formalization and the metaprogramming components of \mathlib.

\subsection{Tactic Database}
\label{subsection:tacdb}

Lean proofs are often developed using tactics.
Custom tactics can be written in the language of Lean as metaprograms,
and \mathlib includes many such tactics~\cite[Sect.\ 6]{mathlib20}.
It is essential for us to provide an index of the available tools
explaining when and how to use them.
Tactic explanations are an example of decentralized documentation.
Their implementations appear in many different files,
interspersed with many other declarations,
but users must see a single unified list.
These same concerns apply to the commands defined in \mathlib,
as well as to attributes and hole commands,
which we do not discuss in this paper.

It is inconvenient to maintain a database of tactics separate from the library.
Since \mathlib changes rapidly,
such a database would likely diverge from the library before long.
In addition,
the doc strings for tactics---which appear as tooltips in supported editors---often
contain the same text as a tactic database entry.
To avoid these issues,
we provide a command \lean{add_tactic_doc}
that registers a new tactic documentation entry.
Another command retrieves all tactic doc entries
that exist in the current environment.

\begin{figure}[t]
\begin{lstlisting}
structure tactic_doc_entry :=
(entry_name               : string)
(category                 : doc_category)
(decl_names               : list name)
(tags                     : list string := [])
(description              : string      := "")
(inherit_description_from : option name := none)

add_tactic_doc
{ entry_name := "linarith",
  category   := doc_cagetory.tactic,
  tags       := ["arithmetic", "decision procedure"],
  decl_names := [`tactic.interactive.linarith] }
\end{lstlisting}

\caption{The information stored in a tactic documentation entry,
and the standard way to register an entry.
The text associated with this entry will be the declaration doc string of \lean{tactic.interactive.linarith}.}
\label{figure:tacdoc}
\end{figure}

A tactic doc entry (Fig. \ref{figure:tacdoc}) contains six fields.
The command \lean{add_tactic_doc} takes this information as input.
To avoid duplicating information,
the \lean{description} field is optional,
as this string has often already been written as a declaration doc string.
When \lean{description} is empty, the command will source it from
the declaration named in \lean{inherit_description_from} (if provided)
or the declaration named in \lean{decl_names} (if this list has exactly one element).
The HTML generation tool links each description to its associated declarations.

The \lean{entry_name} field titles the entry.
This is typically the name of the tactic or command,
and is used as the header of the doc entry.
The \lean{category} field is either \lean{tactic}, \lean{command},
\lean{hole_command}, or \lean{attribut}\lean{e}. 
These categories are displayed on separate pages.
The \lean{decl_names} field lists the declarations associated with this doc entry.
Many entries document only a single tactic,
in which case this list will contain one entry,
the implementation of this tactic.

The \lean{tags} field contains an optional list of tags.
They can be used to filter entries in the generated display.
The command can be called at any point in any Lean file,
but is typically used immediately after a new tactic is defined,
to keep the documentation close to the implementation in the source code.
The HTML display allows the user to filter declarations by tags---e.g.\,
to view only tactics related to arithmetic.

\subsection{Library Notes}
\label{subsection:libnote}

The interface surrounding a definition
is often developed in the same file as that definition.
We typically explain the design decisions of a given module
in the file-level documentation.
However, some design features have a more distributed flavor.
An example is the priority of type class instances (Sect.~\ref{subsection:typeclass}).
There are guidelines for choosing a priority for a new instance,
and an explanation why these guidelines make sense,
but this explanation is not associated with any particular module:
it justifies design decisions made across dozens of files.

We use a mechanism that we call \emph{library notes} (Fig.~\ref{figure:libnote}),
inspired by a technique used in the Glasgow Haskell Compiler~\cite{Marl12} project
to document these distributed design decisions.
A library note is similar to a module doc string,
but it is identified by a name
rather than a file and line.
As with tactic doc entries,
we provide commands in \mathlib
to declare new library notes
and retrieve all existing notes.

\begin{figure}[t]
\begin{lstlisting}
-- declare a library note about instance priority
/-- Certain instances always apply during type class resolution... -/
library_note "lower instance priority"

-- reference a library note in a declaration doc string
/-- see Note [lower instance priority] -/
@[priority 100]
instance t2_space.t1_space [t2_space α] : t1_space α := ...

-- print all existing library notes
run_cmd get_library_notes >>= trace
\end{lstlisting}

\caption{Library notes can be declared, referenced, and collected anywhere in \mathlib.}
\label{figure:libnote}
\end{figure}

The documentation processing tool
generates an HTML page
that displays every library note in \mathlib.
When these notes are referenced
in other documentation entries
with the syntax \lean{Note [note name]},
they are linked to the entry on the notes page.
Library notes are also often referenced in standard comments
that are not displayed in documentation.
These references are useful for library developers
to justify design decisions in places that do not face the public.

\section{Conclusion}
\label{section:conclusion}

Although there are a growing number of large libraries of formal proofs,
both mathematical and otherwise,
little has been written about
best practices for maintaining and documenting these libraries.
Ringer et al~\cite{Ring19} note the gap between proof engineering and software engineering in this respect.
Andronick~\cite{Andr19} describes the large-scale deployment
of the seL4 verified microkernel,
focusing on the social factors that have led to its success;
Bourke et al~\cite{Bour12} describe technical aspects of maintaining this project.
Other discussions of large libraries~\cite{Banc18, Gont13} touch on similar topics.
Wenzel~\cite{Wenz19} explains the infrastructure
underlying the Isabelle Archive of Formal Proofs (AFP),
including progress toward building the AFP with semantic document markup.

Sakaguchi~\cite{Saka20} describes a tool for
checking and validating the hierarchy of mathematical structures
in the Coq Mathematical Components library~\cite{Mahb17},
a task in the same spirit as our type class linters.
Cohen, Sakaguchi, and Tassi~\cite{Cohe20} implement a related tool
which greatly simplifies changing this hierarchy.

It is hard to quantify the effect
that our linters and documentation have had on the \mathlib community.
Fixing issues identified by
the \lean{instance_priority} and \lean{dangerous_instance} linters
led to performance boosts in the library.
Removing unusable instances and simplification lemmas
has also improved performance and decluttered trace output.
More noticeable is the effect on the workload of maintainers,
who can now spend more review time on the deeper parts of library submissions.
Similarly, inexperienced contributors worry less about
introducing subtle mistakes into the library.
Users at all levels report frequent use of the HTML documentation,
especially to find information that is not easily available in an interactive Lean session,
such as the list of instances of a given type class.

So far we have only implemented
the very basic sanity checks on simp lemmas
described in Sect.~\ref{subsection:simp-linters}.
There are also other properties of term rewriting systems
that we want for the simp set,
such as confluence and termination.
Kaliszyk and Sternagel~\cite{KaliszykS13} have used
completion of term rewriting systems
to automatically derive a simp set
for the HOL Light standard library.
We plan to implement a more manual approach,
where a linter detects the lack of local confluence
and prints a list of equations for
the non-joinable critical pairs.
It is then up to the user to decide how to name, orient,
and generalize these new equations.

The current linter framework considers each declaration locally,
but we anticipate the need for global tests.
The \lean{simp_nf} linter
already goes beyond strictly local checking:
it considers the entire simp set.
Another global linter could
check the termination of the simp set.
This is a much harder challenge,
since checking termination is undecidable in general.
We plan to investigate the integration
of external termination checkers
such as AProVE~\cite{Giesl2017}. 

While many of the features we present are specific to Lean,
we believe that the general considerations apply more broadly:
automated validation and documentation seem essential
for a sustainable and scalable library of formal proofs.
Especially in regard to documentation,
there is a definite path for coordination between libraries and systems,
possibly aided by tools from the mathematical knowledge management community.

\paragraph{Acknowledgments.}
We thank Jeremy Avigad and Jasmin Blanchette for comments on a draft of this paper,
and Bryan Gin-ge Chen for many contributions to the \mathlib documentation effort.

%
%
%
 \bibliographystyle{splncs04}
 \bibliography{mathlib-paper}

\end{document}